\documentstyle[twoside,fleqn,espcrc2]{article}

\newcommand{\AmS}{{\protect\the\textfont2
  A\kern-.1667em\lower.5ex\hbox{M}\kern-.125emS}}

\title{ \hfill IASSNS-HEP-93/75\\{\bf Recent developments in chiral gauge
theories:}\\
{\bf Approach of infinitely many fermi fields}}

\author{Rajamani Narayanan\address{School of Natural Sciences, Institute for
        Advanced Study, Olden Lane, Princeton, NJ 08540, USA}
\thanks{Plenary talk presented at the Lattice '93 conference held between
October 12-16 in Dallas, Texas}
\thanks{ This work was supported in part by the DOE grant
\#DE-FG02-90ER40542.}}

\begin{document}

\begin{abstract}
I present the recent developments in a specific sub-field of chiral gauge
theories on the lattice.  This sub-field pertains to the use of infinitely many
fermi
fields to describe a single chiral field. In this approach, both anomalous and
anomaly free theories can be discussed in equal footing. It produces the
correct anomaly in the continuum limit. It has the potential to describe
fermion number violating processes in the presence of a gauge field background
with non-trivial topological charge on a finite lattice.
\end{abstract}

\maketitle

\section{INTRODUCTION}

 Establishing the existence of chiral gauge theories
outside perturbation theory, with fermionic matter in an anomaly-free complex
representation of the gauge group, is a long standing open problem in lattice
field theory.  Many approaches have been proposed and analyzed in the past
and this topic has been an active field of research as is evident from the
review talks in the past lattice conferences \cite{Petcher}. Unfortunately, the
success has
been very limited. The difficulty, as everyone present in the conference knows
very well, is a conceptual one and is not the absence of fast enough computers
\cite{Smit}.

 The conceptual difficulty arises from the presence of anomalies in a chiral
gauge theory. This makes the regularization a very subtle problem. This is
a well understood issue in the realm of perturbation theory. There has also
been considerable progress in the past to elevate the continuum regularization
to a non-perturbative level. This has been very useful in bringing out the
subtleties present in chiral gauge theories. But a lattice regularization of
chiral gauge theories is needed to provide a mundane framework for definite
computations. It is needed to understand the continuum limit of chiral gauge
theories, particularly the difference between anomalous and anomaly free
theories. It is needed to non-perturbatively compute `t Hooft processes
\cite{`t Hooft}.

 I shall deviate from the conventional line of thought right from the start.
The difficulties encountered by the previous approaches will greatly influence
the nature of questions that I address here. The hope is that this
review will complement previous reviews that dealt with chiral fermions and
facilitate future progress in the understanding of chiral theories.  The line
of thought is strongly influenced by two papers: one by Kaplan \cite{Kaplan}
which many of
you are probably familiar with; and the second, a lesser known paper by
Frolov and Slavnov \cite{Fronov}. Several papers have been written on the
pros and cons of Kaplan's proposal in the last one year
\cite{Narger1,Narger2,Narger3,Shamir,Korthals,Aokih,Finite}. I shall try my
best to combine most of the ideas in these papers to present a coherent view of
the new proposal to tackle chiral gauge theories.

The organization of the talk is as follows. I will start by showing how one can
use an infinite set of fermi fields to describe a single chiral fermion in the
continuum \cite{Fronov}. I will then show that this can be extended to the
lattice. The resulting model will be different from the one proposed by Kaplan
\cite{Kaplan} in that the gauge fields will be four dimensional and not five
dimensional. I will briefly discuss the difficulties in working with five
dimensional gauge fields. The need for infinitely many fermi fields seems to
make the model impractical at first sight. I will show that some attempts to
use
a finite set of fields destroys the chiral nature of the theory and makes it
vector-like \cite{Finite}.
Next,
I will show how one can deal with an infinite number of fermi fields in a
practical calculation without restricting them to a finite set. I will focus
on the effective action induced by fermions and will show that it is an overlap
of two second quantized ground states. This provides a non-perturbative formula
for the effective action on a finite lattice whose merits can be explicitly
tested. At this point, I will make a formal extension of the overlap to the
continuum and show how this formula is a definition of the "chiral determinant"
in the continuum. Since the effective action is an overlap of two different
states, there is an ambiguity in its imaginary part. I will show that
perturbation theory provides a way of fixing this ambiguity.
Finally, I will show that because the effective action
is an overlap of two ground states, it has the potential to vanish exactly
for some gauge backgrounds. Further, insertion of an appropriate number of
fermion operators in between the two ground states will make it non-zero in
such gauge backgrounds. In this way, I will obtain the `t Hooft picture
\cite{`t Hooft} on the
lattice and also get a definition for the topological charge that is always
an integer.

\section{INFINITELY MANY FERMI FIELDS}

Consider the Pauli Villars regularization of a single chiral fermion
transforming under a complex representation of the gauge group G. A general,
local,
hermitian Lagrangian for the fermions is
\begin{eqnarray}
{\cal L}= & & i\bar\psi_s \gamma^\mu (\partial_\mu - igA_\mu)\psi_s\nonumber\\
& + & \bar\psi_s
(M_{ss'}P_++{M^\dagger}_{ss'} P_-)\psi_{s'}
\end{eqnarray}
$\psi_s$ is a set of $s$ Dirac fields which includes the
original chiral fermion and as many regulator fields as is
needed. $P_{\pm}={1\over 2}(1\pm \gamma^5)$ are the projectors into right and
left handed fields respectively. $A_\mu=A_\mu^aT^a$ with $T^a$ being the
group generators in some complex representation. $M$ is some mass matrix. The
structure,
$\gamma_0(MP_++M^\dagger P_-)$ is hermitian. To describe a single right handed
chiral
fermion, $M$ should have one zero mode and $M^\dagger$ should not have any,
i.e, $M$ should have an index equal to unity.
This cannot be achieved by a finite dimensional matrix $M$. Restriction to
a finite set of fields will force the chiral fermions to occur in pairs of
opposite chirality.
If $M$ is infinite dimensional, one can construct
an $M$ with non-zero index. A simple
example is to choose $M$ proportional to the annihilation operator of an
harmonic
oscillator where $s$ plays the role of the label associated with the energy
levels of the harmonic oscillator.  All the heavy modes of $M$ and all the
modes of $M^\dagger$ correspond to regulator fields and should be taken to
infinity.
One can now proceed to see if the statistics of the infinitely many fields can
be
suitably chosen to obtain a well regulated theory. It can be shown that we
will end up with a well regulated theory (in the perturbative sense) if the
original theory is anomaly free \cite{Fronov,Narger1}.

To extend the above idea to a non-perturbative level, consider the lattice
regularization of the same problem. One can start with a Langrangian similar to
(1) and extend it to the lattice in the usual manner.  Now, $M$
should satisfy the following properties. In the region enclosing $p=0$ in
the Brillouin zone, $M$ should have an index equal to unity.
But the zero mode of $M$ should disappear as one goes near the
edges of the Brillouin zone where the potential doublers exist. One can achieve
such an $M$  by considering an infinite set
of Wilson fermions with a $p$ independent mass matrix coupling them. A specific
action for fermions that achieves this goal is
\begin{eqnarray}
S_F(\bar\psi, \psi, U) &=& \ \ {1\over 2} \sum_{n,s,\mu}
\bar\psi_{n,s}(1+\gamma^\mu)U_{n,\mu}\psi_{n+\hat\mu,s}\nonumber \\
& & +{1\over 2} \sum_{n,s,\mu}
\bar\psi_{n,s}(1-\gamma^\mu)U^\dagger_{n-\hat\mu,\mu}\psi_{n-\hat\mu,s}\nonumber\\
& & +{1\over 2} \sum_{n,s} \bar\psi_{n,s}(1+\gamma^5)\psi_{n,s+1}\\
& & +{1\over 2} \sum_{n,s} \bar\psi_{n,s}(1-\gamma^5)\psi_{n,s-1}\nonumber\\
& & -\sum_{n,s} \Bigl[5-m\ {\rm sign} (s+{1\over
2})\Bigr]\bar\psi_{n,s}\psi_{n,s}\nonumber
\end{eqnarray}
$\bar\psi_{n,s}$ and $\psi_{n,s}$ are Dirac spinors. $m$ is a parameter in the
region $0< m < 1$.
$n$ is a four component integer labeling sites on the lattice and $s$ labels
the infinite copies of fermions. $n$ can run over a finite set or
an infinite set of integers. When it runs over a finite set, which will be the
case of
practical interest, the boundary conditions on the fermions will be chosen to
be anti-periodic. The above action will be referred to as the ``wall"
realization of a single chiral fermion. Although,
this is basically the idea of Kaplan \cite{Kaplan}, the line of thought
presented
here is significantly different from his and the action itself has some
differences.

Kaplan \cite{Kaplan} viewed the infinite set
of fermi fields as residing in a fifth direction. As such, he was originally
inclined
to couple then to a five dimensional gauge field in the spirit of Callan and
Harvey \cite{Calvey}.
This introduces a lot of unwanted degrees of freedom. Wild fluctuations of the
gauge field in the fifth direction could essentially decouple adjacent $s$
slices and spoil the chiral zero mode. It is shown in the context of mean field
analysis that increasing the coupling in the fifth direction brings about a
layered phase \cite{Korthals,Finite}. Perturbative analysis of a 2+1 (two
physical
and one $s$ direction) model with three dimensional  U(1) gauge
fields has been done \cite{Aokih}. Here
one obtains the parity-odd terms that break 2-D gauge invariance but one also
gets a longitudinal piece in the real part that does not vanish even for
anomaly free cases. These investigations seem to indicate that working with
higher dimensional gauge fields is not a good direction to proceed.
In (2) all the fermi fields are coupled to the same gauge field. This is in the
same spirit as in Pauli-Villars regularization. The gauge fields are therefore
the usual four dimensional fields.

Kaplan, in a later proposal \cite{Kaplan1}, did suggest a model with four
dimensional gauge fields. But he made the fifth direction finite because
having an infinite direction seems to make a numerical simulation impossible at
first sight. Making the fifth direction finite results in the theory having
two chiral modes of opposite chirality. This is called the ``wall--antiwall"
model, where one zero mode is centered on the wall and the other zero mode is
centered on the anti-wall. To decouple one of the zero modes from the gauge
field, Kaplan decided to put in gauge fields only on the slices around the
wall and not have gauge fields near the anti-wall. The hope was that if the
wall and anti-wall were far removed from each other the resulting theory
in the vicinity of the wall would be a chiral gauge theory. Gauge invariance
is broken in this model and was restored by the addition of scalar fields
on the two $s$ slices where the gauge field is cut-off.
This model was analyzed in some detail by Golterman et.al. \cite{Finite}.
The introduction of the scalar fields to restore gauge invariance introduces
Yukawa coupling. For zero Yukawa coupling they found two more massless modes
centered on the slices where the scalar field is introduced. A search for
a region of the parameter space where the scalar field and the extra zero modes
are all decoupled from the physical zero mode was not met with success. There
is no obvious reason why this model should not succeed nor was the search
performed in \cite{Finite} exhaustive enough to rule out this approach.
In the next section, I will show that it is possible to deal with an infinite
set of fields directly and still be able to deal with the problem numerically
in a computer
with finite memory size in a finite amount of time!

There are two other technical differences between
(2) and Kaplan's proposal. One is that the Wilson coefficient is set to
$r=0.5$. This is to ensure positivity in the transfer matrix formalism. The
other difference is that the defect is between $s=0$ and $s=-1$. This
simplifies the result obtained from the transfer matrix formalism.

\section{THE OVERLAP FORMULA}

In this section, I will discuss the implementation of the "wall" realization
on a finite lattice by keeping the number of fermi fields infinite. I will
concentrate only on the effective action induced by the fermions and
present the arguments and the results. For derivations, the reader is
referred to \cite{Narger2}.

Consider the action in (2) on a $L^4\times \infty$ lattice. The lattice is
finite in the four spatial directions and is infinite in the $s$ "direction".
In each slice $s$ there are a finite number of degrees of freedom, both
fermionic
and bosonic. Further, the gauge fields are the same in all slices. The
transfer matrix is obtained by the standard translation of the
Grassmann integration into the operator formalism. We will choose the
propagation along the $s$ direction and the transfer matrix will connect
two adjacent $s$ slices. Since the action in (2) is uniform for $s<0$ and
for $s\ge 1$, the transfer matrices connecting slices $s$ for $s<0$ will
all be the same and so will the transfer matrices connecting slices $s$ for
$s\ge 1$. This implies that there are only two finite
dimensional transfer matrices in the problem. The result for the effective
action, $S_{\rm eff}(U)$, is
\begin{eqnarray}
e^{S_{\rm eff}(U)} =\lim_{s\rightarrow\infty}&&
[\det {\bf B}^-]^{s+{3\over 2}} [\det {\bf B}^+]^{s+{3\over 2}}\nonumber\\
&&<b-|e^{-s\hat H_-} e^{-s\hat H_+}|b+>\nonumber
\end{eqnarray}
\[
\hat H_\pm=-\hat a^\dagger {\bf H}_\pm \hat a;
\]
\begin{equation}
e^{{\bf H}_\pm}=\pmatrix{{1\over {\bf B}^\pm} & {1\over {\bf B}^\pm}{\bf C}\cr
{\bf C}^\dagger {1\over {\bf B}^\pm} &
 {\bf C}^\dagger {1\over {\bf B}^\pm} {\bf C} + {\bf B}^\pm \cr}
\end{equation}
\begin{eqnarray}
{\bf B}^\pm_{n\alpha i,m\beta j}&&=
 (5\mp m)\delta_{nm}\delta_{\alpha\beta}\delta_{ij}\nonumber\\
&&-{1\over 2}\delta_{\alpha\beta}\sum_\mu\Bigl[
 \delta_{m,n+\hat\mu}U^{ij}_{n,\mu}
 +\delta_{n,m+\hat\mu}{U^{ji}_{m,\mu}}^*\Bigr]\nonumber
\end{eqnarray}
\[
 {\bf C}_{n\alpha i,m\beta j}=
 {1\over 2}\sum_\mu\Bigl[
 \delta_{m,n+\hat\mu}U^{ij}_{n,\mu}
 -\delta_{n,m+\hat\mu}{U^{ji}_{m,\mu}}^*\Bigr]
 \sigma_\mu^{\alpha\beta}
\]
 $\hat a^\dagger$ and $\hat a$ are fermion operators satisfying canonical
 anti-commutation relations.
\[
 \{\hat a_{nAi}, \hat a^\dagger_{mBj}\}=\delta_{nm}\delta_{AB}\delta_{ij};
\]
\begin{equation}
 \{\hat a_{nAi}, \hat a_{mBj}\}=
 \{\hat a^\dagger_{nAi}, \hat a^\dagger_{mBj}\}=0.
\end{equation}
 The indices $A$ and $B$ run over the four spinor components whereas the
 indices $\alpha$ and $\beta$ run over two spinor components.
 The $\gamma$ matrices are
 $\gamma^\mu=\pmatrix {0 & \sigma_\mu\cr \sigma^\dagger_\mu & 0\cr}$;
 $\gamma^5=\pmatrix {1 & 0\cr 0 & 1\cr}$, where $\sigma_0=i$ and $\sigma_k$;
 $k=1,2,3$ are the usual Pauli matrices. ${\bf B}^\pm$ are positive definite
 for all values of the gauge fields. ${\bf H}_\pm$ are both hermitian and
 traceless. $|b\pm>$ are the boundary conditions at $s=\pm\infty$.

 As it stands, the effective action in (3) is not finite. This is due to
two contributions. One comes from the two determinants $\det {\bf B}^\pm$
raised to infinite power. As $s\rightarrow\infty$, $e^{-s\hat H_\pm}$ go into
the respective projectors $e^{-(\lambda^0_\pm)s}|0\pm><0\pm|$ where $|0\pm>$
and $\lambda^0_\pm$ are the ground state eigenvectors and eigenvalues
of $\hat H_\pm$ respectively. The second infinity comes from
$e^{-(\lambda^0_\pm)s}$ as $s\rightarrow\infty$. These are both bulk type
infinites and to remove them
consider the effective actions, $S^\pm_{\rm eff}(U)$,
due to the homogeneous cases obtained by replacing the coefficient of the mass
term in the action (last term in (2)) by $5\mp m$ respectively.
\begin{equation}
e^{S^\pm_{\rm eff}(U)} =\lim_{s\rightarrow\infty}
[\det {\bf B}^\pm]^{2s+3} <b\pm|e^{-2s\hat H_\pm}|b\pm>
\end{equation}
Note that $S^\pm_{\rm eff}(U)$ are both real. To cancel the bulk infinities,
we define the interface effective action as
\begin{equation}
S_i(U)=S_{\rm eff}(U)-{1\over 2}[S^+_{\rm eff}(U)+S^-_{\rm eff}(U)]
\end{equation}
{}From (3) and (5), as $s\rightarrow\infty$,
\begin{equation}
e^{S_i(U)}={<b-|0-><0-|0+><0+|b+>\over |<b-|0->||<b+|0+>|},
\end{equation}
and $S_i(U)$ is finite. Note that the boundary states at $\pm\infty$ in
(3) and (5) were appropriately chosen for the inhomogeneous and homogeneous
cases. Choosing periodic boundary conditions in the $s$ direction
amounts to setting $|b+>=|b->$ and summing over a complete set of states.
This will make the right hand side of (7) equal to $|<0-|0+>|^2$. It is
real and corresponds to an effective action coming from two chiral modes of
opposite chirality and is precisely what has to be avoided. The two ends,
$s\rightarrow\pm\infty$, should be kept separate. The appropriate boundary
states are $|b\pm>=|0\pm>$. This is precisely what one would pick if one is
thinking of a quantum mechanical path integral. The result is the overlap
formula for the effective action:
\begin{equation}
$$e^{S_i(U)}=<0-|0+>
\end{equation}
The above formula is finite and non-perturbative.

\section{CONTINUUM INTERPRETATION}

I will now show that the simple formula, (8), obtained via a circuitous route
starting from infinitely many fermi fields, has a simple continuum
interpretation. The contents of this section are in no way rigorous. Consider
the following Hamiltonian in the continuum:
\begin{equation}
-{\bf H}=\pmatrix{m & X^\dagger \cr X & -m }.
\end{equation}
$X$ is the chiral Dirac operator and ${\bf H}$ with $m>0$ and $m<0$ are the
formal continuum limits of the Hamiltonians discussed in the previous section.
The states $|0\pm>$ that enter the continuum limit of (8) are the ground states
obtained by filling all the negative energy states
of ${\bf H}$ with $m>0$ and $m<0$ respectively. Let
\begin{equation}
-{\bf H}\psi_\lambda
=\lambda\psi_\lambda;~~~~~~~\psi_\lambda=\pmatrix{u_\lambda
\cr v_\lambda}
\end{equation}
be the eigenvalue equation for ${\bf H}$.  Then
\begin{equation}
X^\dagger v_\lambda =  ( \lambda - m) u_\lambda;\ \ \
XX^\dagger v_\lambda = (\lambda^2 -m^2 ) v_\lambda
\end{equation}
\begin{equation}
X u_\lambda = ( \lambda +m ) v_\lambda;\ \ \
X^\dagger X u_\lambda = (\lambda^2 -m^2 ) u_\lambda
\end{equation}
For $m=|m|$, the complete set of negative energy solutions are given by
\begin{equation}
\psi_i^+ =N_i^+ \pmatrix{ u^{(i)} \cr X{1\over {\sqrt {X^\dagger X + m^2}
+|m|}}u^{(i)}}
\end{equation}
where we used (12). For $m=-|m|$, the complete set of negative energy solutions
are given by
\begin{equation}
\psi_i^- =N_i^- \pmatrix{  X^\dagger {1\over {\sqrt {XX^\dagger + m^2} +|m|}}
v^{(i)} \cr v^{(i)}}
\end{equation}
where we used (11).Note that both (13) and (14) are well defined even for the
zero modes of $XX^\dagger$ or $X^\dagger X$. The superscript $i$ denotes the
index of the eigenvector and $N_i^\pm$ are the normalization factors. The
overlap is then given by
\begin{eqnarray}
<0-|0+>&&=
\det_{ij} \lbrack 2N_i^- N_j^+\\
 &&< v^{(i)} , {1\over {\sqrt {XX^\dagger + m^2}
+|m|}} X u^{(j)} >\rbrack\nonumber
\end{eqnarray}
This expression can be interpreted as a "determinant" of $X$. $\{u^{(i)}\}$ is
a basis set that spans the space on which $X$ acts. The action of $X$ takes it
to another space for which the set $\{v^{(i)}\}$ is a basis. The inner product
on the right hand side of (15) is a matrix element of $X$. The expression (15)
is (partially) regularized and its phase will be the imaginary part of
the effective action.

\section{THE PHASE AMBIGUITY}

(8) still has an ambiguity arising from the phase choice of the two ground
states $|0\pm>$. Different choice of phases affect the imaginary part of the
effective action wherein lies the anomaly content of the theory. In particular,
it is not clear why the phases cannot be chosen to make the overlap purely
real. This issue can be resolved by the use of continuum perturbation theory
\cite{Narger2}. Let
\[
\Psi^+_p = \psi^+_p + \psi^+_q A_{qp} + \chi^+_q B_{qp}
+ \psi^+_q W_{qp} + \chi^+_q X_{qp}\]
\begin{equation}
\Psi^-_p  \psi^-_p + \psi^-_q C_{qp} + \chi^-_q D_{qp}
+ \psi^-_q Y_{qp} + \chi^-_q Z_{qp}.
\end{equation}
$\psi^\pm_p$ are eigenvectors corresponding to negative eigenvalues and
$\chi^\pm_p$ are eigenvectors corresponding to positive eigenvalues of
the gauge free Hamiltonians, ${\bf H}_\pm(0)$.
$A,B,C$ and $D$ are coefficients linear in the gauge field and $W,X,Y$ and $Z$
are coefficients quadratic in the gauge field.
Perturbation theory does not fix the diagonal terms of
$A$, $W$, $C$ and $Y$. Normalization fixes the real part but the imaginary part
remains ambiguous. Any choice to fix this ambiguity has to be local functions
of the gauge fields.
 Computation of $S_i(A_\mu)$ in 2-D with U(1) gauge fields, $A_\mu$, gives
\[{\rm Im}[S_i(A)-S_i(0)]= I_1+I_2+I_3\]
\[I_1={1\over 2} {\rm Tr}(A-A^\dagger+C^\dagger-C)\]
\begin{equation}
I_2={1\over 2} {\rm Tr}(W-W^\dagger+Y^\dagger-Y)
\end{equation}
\[I_3={1\over 16\pi} \sum_p\Bigl[
{p_1+ip_2\over p_1-ip_2}z^*_pz^*_{-p}
-{p_1-ip_2\over p_1+ip_2}z_pz_{-p}\Bigr]\]
\[z_p=\int d^2x e^{ipx} [A_1(x) + i A_2(x)]\]
$I_1$ and $I_2$  are completely ambiguous. $I_3$ is completely fixed and
nonlocal. To make ${\rm Im}[S_i(A)-S_i(0)]=0$, the diagonal terms of $A$, $W$,
$C$ and $Y$ have to be chosen so that $I_1=0$ and $I_2$ cancels $I_3$.
This cannot be achieved due to the nonlocality of $I_3$.
 The anomaly is a consequence of the non-local nature of $I_3$.
 Any choice for the diagonal terms in $W$ and $Y$, that is local, is a good
choice and will not affect the anomaly.

One choice is Brillouin-Wigner perturbation theory where
${{}_{U}\negthinspace\negthinspace<0\pm |0\pm >_{1}}$
is chosen to be real. The resulting formula for the effective
action is
\[
e^{S_i (U)} ={
{{}_{U}\negthinspace\negthinspace<0-|0+>_{U}}
\over {{}_{1}
\negthinspace\negthinspace<0-|0+>_{1}}}
e^{i[\Phi_+ (U) - \Phi_- (U)] }
\]
\begin{equation}
e^{i\Phi_\pm (U) }= {{{}_{U}\negthinspace\negthinspace<0\pm |0\pm >_{1}}
\over {| {{}_{U}\negthinspace\negthinspace<0\pm |0\pm >_{1}
}|}}
\end{equation}
This formula was shown to reproduce the correct continuum anomaly in
\cite{Narger3} by computing (18) on several finite lattices for some fixed
gauge fields and taking the continuum limit.

(18) has a very interesting property of becoming zero for some gauge fields.
 This can happen as follows. In a finite lattice ${\bf H}_\pm$ are both finite
dimensional matrices. The ground state is obtained by filling all the negative
energy states of ${\bf H}_\pm$. The overlap of the two ground states will be
non-zero only if the number of negative energy states of ${\bf H}_\pm$ are the
same. In the gauge free case, half of the states are negative for both ${\bf
H}_\pm$. This situation is expected to be true for perturbative gauge fields.
Here the overlap will be non-zero. One can come up with gauge field
configurations where this is not the case \cite{Narger3}. Fluctuations around
these gauge configurations cannot change this property. Since the overlap is
zero due to a mismatch between the number of negative energy states of
${\bf H}_\pm$, it is easy to see that insertion of an appropriate number of
fermion operators between the two ground states will make it non-zero.
Therefore, the effective action, (18), reproduces the `t Hooft picture \cite{`t
Hooft} of
the chiral determinant.
It can be proven that ${\bf H}_-$ always has an equal number of positive
and negative eigenvalues for all gauge fields \cite{Narger3}. Only
${\bf H}_+$ can have different number of positive and negative eigenvalues.
Therefore (18) is well-defined for a generic gauge field configuration.

The success of this apporach to chiral fermions depends upon how well (18)
reproduces all the physical results. The interesting point about (18) is that
it interlocks two important features of chiral gauge theories: anomalies and
fermion number violation. Since it is an overlap of two different states, it
can have a phase and reproduce the anomaly \cite{Narger3}. Since it is an
overlap of two different second quantized states, it can be exactly zero for
some gauge configurations \cite{Narger3}. It is relatively easy to reproduce
the anomaly in a conventional approach to chiral gauge theories but to
reproduce fermion number violating processes on the lattice is hard because the
formalism has to be valid for rough gauge fields \cite{Smit1}.

\noindent{\bf Acknowledgements:} I would like to thank Herbert Neuberger for a
joyous and fruitful collaboration \cite{Narger1,Narger2,Narger3}. I would also
like to thank Sinya Aoki, Mike Creutz, Maarten Golterman, Jan Smit, Pierre van
Baal, and Jeroen Vink for several interesting discussions.

\end{document}